\documentstyle[preprint,epsf,aps ]{revtex}
\draft

\def\beq{\begin{equation}}
\def\eeq{\end{equation}}

\def\prb{Phys. Rev. B }
\def\pra{Phys. Rev. A }
\def\pla{Phys. Lett. A }
\def\pt{Physics Today}
\def\sm{Superlattices and Microstructures}
\def\prl{Phys. Rev. Lett. }
\def\ajp{Am. J. Phys. }

\def\mplb{Mod. Phys. Lett. B }

\def\ibmjrd{IBM J. Res. Dev. }
\def\pjp{Pramana J. Phys. }

\begin{document}
\draft
\title{Effect of quantum entanglement on  Aharonov-Bohm oscillations,
  spin-polarized transport and current magnification effect}
\author{A. M. Jayannavar\cite{amje} }
\address{Institute of Physics, Sachivalaya Marg, Bhubaneswar 751 005,
  Orissa, India}
\maketitle

\begin{abstract}

  We present a simple model of  transmission across a metallic
  mesoscopic ring. In one of its arm an electron interacts with a
  single magnetic impurity via  an exchange coupling. We show that
  entanglement between electron and spin impurity states leads to
  reduction of Aharonov-Bohm oscillations in the transmission coefficient.
  The spin-conductance is asymmetric
  in the flux reversal as opposed to the two probe electrical
  conductance which is symmetric. In the same model in contradiction
  to the naive expectation of a current magnification effect, we
  observe enhancement as well as the suppression of this effect
  depending on the system parameters. The limitations of this model to 
  the general notion of dephasing or decoherence in quantum systems
  are pointed out.

\vskip 2.0cm

{\bf keywords} Entanglement, Aharonov-Bohm effect, spin-conductance, 
dephasing, decoherence, current magnification.
\vskip 2.0cm

{\bf PACS Nos  73.23.-b, 05.60.Gg, 72.10.-d, 03.65.Bz}

\end{abstract}
\newpage
\section{Introduction} 
In recent times there is a great deal of interest in mesoscopic
systems, sparked by the advancement of technology. Experimental
investigations on these systems have provided several surprising
quantum behavior in total contrast to that anticipated from the
classical theory of metals. One of the prominent mesoscopic effects is
that of Aharonov-Bohm(AB) oscillations in the transport property of
normal metal rings enclosing magnetic
flux\cite{dephase:imry,dephase:datta,dephase:psd,dephase:webb_ap,dephase:gia,dephase:wgtr}.
Here AB oscillations are revealed \cite{dephase:webb_ap} in the
resistance of a small metal ring as a function of the magnetic field
with a period equal to $\phi_{0}=hc/e$, the fundamental flux quanta.
The oscillations in the resistance arise from the interference of
electronic waves traversing the two alternative arms of the ring. The
changing magnetic flux alters the relative phase difference between
the probability amplitudes associated with different paths(upper and
lower arms of the loop). The amount of flux $\phi_{0}$ is required to
enforce a $2\pi$ relative phase shift between two alternative paths,
This leads to the constructive and
destructive interference in the transmission of an electron across the
conductor as one tunes the magnetic flux. At  high temperatures
the inelastic scattering length is much larger than the sample
dimensions and as a result the transport is completely phase coherent,
i.e., it is dominated by quantum interference effects. At very low
temperatures the inelastic scattering length is much smaller than the
sample dimensions which leads to classical behavior(loss of
interference). This process is referred to as dephasing or decoherence
as a result of the randomizing of the interfering particle's phase.
The decoherence mechanism signals the limits beyond which the system
dynamics approaches the classical behavior and arises due to the
coupling of a particle to its environment. This subject of intrinsic
decoherence and dephasing is being pursued actively in the area of
mesoscopic physics
  
  In a double slit setup, interference results from the lack of
  knowledge of (or indistinguishability of) the electron path. Thus a
  measurement of which path the electron has taken, wipes out the
  interference pattern.  It is known that in a ring interferometer the
  electron affects the environment and changes its state differently
  in the two arms of the ring thereby affecting the interference. This
  amounts to a measurement of the path of the interfering particle by
  the environment resulting in loss of interference. Such
  interferometers are thus also termed as ``which-path'' detectors. In
  an alternate picture, the environment affects the electron phase
  differently in the two arms, thus randomizing their relative phase
  difference leading to dephasing. The two views were shown to be
  equivalent\cite{dephase:sai}.  It is well known that the
  electron-environment entanglement can also lead to
  decoherence\cite{dephase:schul}. However, unlike other approaches,
  entanglement leads to decoherence even in absence of any energy
  transfer \cite{dephase:sai}. Experiments have been carried out which
  are aimed at measuring these coherence properties and it has been
  observed for instance, by  placing a micro-detector near one arm
  of the AB interferometer causes decoherence\cite{dephase:sprinzak}.
  Thus motivated, we consider a simple model of dephasing in an
  Aharonov-Bohm ring with a spin-half impurity (spin-flipper) in one
  arm. This example also serves to illustrate the effect of multiple
  reflections on "which-path" detection.

  By introducing a magnetic impurity atom (to be referred to as the
  spin-flipper, or the flipper, for short) in one arm of the ring, one
  can couple the spin of the electron ($\vec{\sigma}$) to the spin of
  the flipper ($\vec{S}$) via the exchange interaction
  \cite{dephase:sai,dephase:imry}. This leads to scattering of the
  electron in which the spin state of the electron and the impurity is
  changed without any exchange of energy.  Additionally, this
  scattering leads to the entanglement-induced reduction of
  interference pattern \cite{dephase:schul}. Let the electron be
  incident from the left reservoir with its spin pointing ``up'' (see
  Fig.~\ref{dephase:ring}).  The spin of the electron passing through
  the upper arm may or may not be flipped by the flipper. In the case
  that the spin is unflipped, one would expect the usual
  AB-oscillations of the transmission due to interference of the
  partial waves passing through the upper and the lower branches of
  the ring. However, in the case that the spin is flipped, one would
  think, guided by naive intuition, that a path detection has taken
  place and hence one would be led to conclude that the interference
  pattern for the spin-down component would be wiped out. This is true
  provided we consider only two forward propagating partial waves.
  However, there are infinitely many partial waves in this geometry
  which are to be superposed to get the total transmission. These
  arise due to the multiple reflections from the junctions and the
  impurity site. Consider, for example, an incident spin-up particle
  moving in the upper arm which is flipped at the impurity site and
  gets reflected to finally traverse the lower arm before being
  transmitted. Naturally, this partial wave will interfere with the
  spin-flipped component transmitted along the upper arm. This results
  in non-zero transmission for the spin-flipped electron.  Thus on
  taking into account the multiple reflections (more than just two
  partial waves) the presence of magnetic impurity does not lead to
  "which-path" information.  However, we show that the presence of
  magnetic impurity does lead to the reduction of AB-oscillations.
  
  Within the same model we also study spin-polarized
  transport\cite{dephase:prinz}. We have discussed the symmetry
  properties of reflection and transmission coefficients of different
  spin channels in the presence of magnetic flux.  In particular the
  spin-conductance which is related to the spin polarised transmission
  coefficient is shown to be asymmetric in flux reversal. We also
  study the current magnification
  effect\cite{dephase:cme1,dephase:cme2,dephase:colin}.  In the case
  of a mesoscopic loop with unequal arms connected to two electron
  reservoirs at chemical potentials $\mu_1$ and $\mu_2$ via ideal
  leads, currents $I_{1}$ and $I_{2}$ flow in the lower and upper arm
  respectively of the loop such that total current $I=I_{1}+I_{2}$ is
  conserved in accordance with Kirchoff's law. In general these two
  currents differ in magnitude and are individually smaller than the
  total current $I$. However, in certain range of Fermi energies the
  current $I_{1}$ or $I_{2}$ may become larger than the total current
  $I$. The property that current in one of the arms is larger than the
  transport current is referred to as current magnification effect. To
  conserve the total current at the junctions, the current in the
  other arm becomes negative, i.e., flows against the applied external
  field.  In such a situation one can interpret that the negative
  current flowing in one arm continues to flow as a circulating
  current in the loop. The magnitude of the negative current in one of
  the arms flowing against the direction of the applied current is
  taken to be that of the circulating current.  When the negative
  current flows in the upper arm the circulating current direction is
  taken to be anticlockwise (or negative) and when it flows in the
  lower arm the circulating current direction is taken to be clockwise
  (or positive). The circulating current here arises in the absence of
  magnetic field. Like AB effect, this effect too is purely quantum
  mechanical in origin.  Even though quantum entanglement dephases AB
  oscillations, we find however, that in contradiction to the naive
  expectation of a reduction of current magnification, it leads to
  enhancement as well as suppression of the effect.  This fact points
  out the limitations of a model based on the interaction induced
  entanglement of quantum states to the general understanding of
  dephasing in quantum systems.

\section{Theoretical Treatment}
\begin{figure*}
\protect\centerline{\epsfxsize=4in \epsfbox{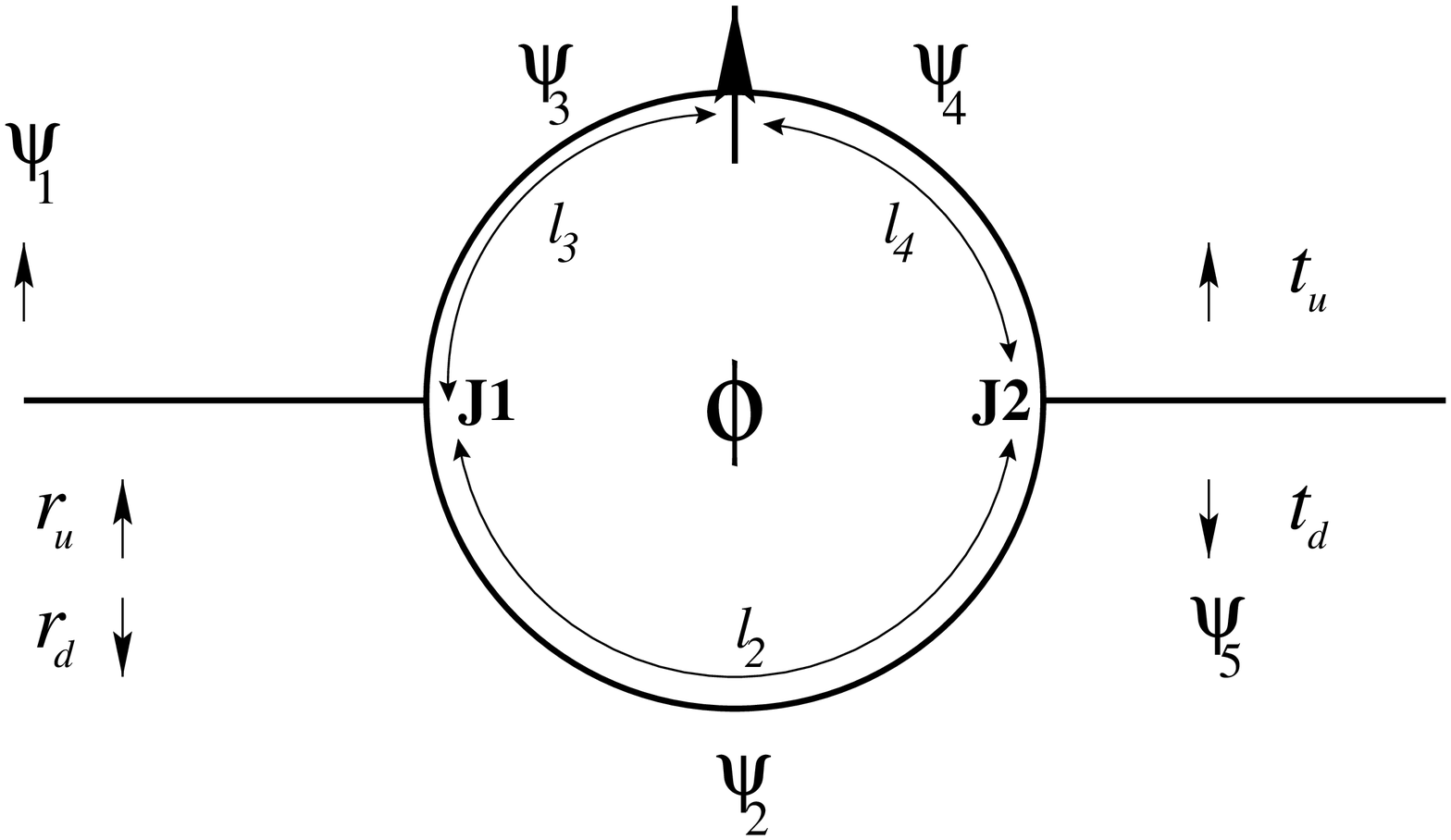}}
\caption{Mesoscopic ring with Aharonov-Bohm flux $\phi$ threading
  through the center of the ring and a magnetic impurity in one arm of
  the ring.}
\label{dephase:ring}
\end{figure*}

We study the problem using the quantum waveguide theory approach
\cite{dephase:wgtr,dephase:cme1,dephase:cme2,dephase:xia} and the spin
degree of freedom of the electron is dealt with in line with Ref.
\onlinecite{dephase:ajp}. We consider an impurity consisting of a
flipper capable of existing in M different discrete internal spin
states and located at a particular position on the upper arm of the
ring (see Fig. \ref{dephase:ring}). The spin $\vec{\sigma}$ of the
electron couples to the flipper spin $\vec{S}$ via an exchange
interaction $-J \vec{\sigma} \cdot \vec{S} \delta(x-l_3)$. The
magnetic flux threading the ring is denoted by $\phi$ and is related
to the vector potential $A=\phi/L$, $L$ being the ring circumference
\cite{dephase:xia}.  During passage of the electron through the ring,
the total spin angular momentum and its $z$-component remain
conserved. We consider the incident electron to be spin-polarized in
the up-direction.

Let $l_2$ be the length of the lower arm of the ring and the impurity
atom be placed at a distance $l_3$ from the junction J1, $l_4$ being
the remaining segment length of the upper arm. The various segments of
the ring and its leads are labeled as shown in Fig. \ref{dephase:ring}
and the wave functions in these segments carry the corresponding
subscripts. The wave functions in the five segments for a
left-incident spin-up electron can be written as
follows\cite{dephase:wgtr,dephase:xia,dephase:ajp}:

\begin{eqnarray}
\label{dephase:eq:1}
\psi _1&=&(e^{ikx}+r_u e^{-ikx})\chi _m\alpha+r_d e^{-ikx}\chi _{m+1}\beta,\nonumber\\
\psi _2&=&(A_u e^{ik_1x}+B_u e^{-ik_2x})\chi _m\alpha+(A_d e^{ik_1x}+B_d e^{-ik_2x})\chi _{m+1}\beta,\nonumber\\
\psi _3&=&(C_u e^{ik_1x}+D_u e^{-ik_2x})\chi _m\alpha+(C_d e^{ik_1x}+D_d e^{-ik_2x})\chi _{m+1}\beta,\nonumber\\
\psi _4&=&(E_u e^{ik_1x}+F_u e^{-ik_2x})\chi _m\alpha+(E_d e^{ik_1x}+F_d e^{-ik_2x})\chi _{m+1}\beta,\nonumber\\
\psi _5&=&t_u e^{ikx}\chi _m\alpha+t_d e^{ikx}\chi _{m+1}\beta.
\end{eqnarray}

\noindent where $k_1=k+(e\phi /\hbar cL)$, $k_2=k-(e\phi /\hbar cL)$,
$k$ is the wave-vector of incident electron. The wavefunction in
Eqn.(1) is a correlated function (entangled state) of
the electron and the impurity spin which takes into account that the
exchange interaction conserves the $z$-component of the total
spin\cite{dephase:ajp}.  The subscripts $u$ and $d$ represent ``up''
and ``down'' spin states of the electron with the corresponding
spinors $\alpha$ and $\beta$ respectively (i.e., $\sigma _z\alpha
=\frac{1}{2}\alpha$, $\sigma _z\beta =-\frac{1}{2}\beta$) and $\chi
_m$ denotes the wave function of the impurity \cite{dephase:ajp} with
$S_z=m$ (i.e., $S_z\chi _m=m\chi _m$).  The reflected (transmitted)
waves have amplitudes $r_u$ ($t_u$) and $r_d$ ($t_d$) corresponding to
the ``up'' and ``down'' spin components respectively.

\noindent  Equations (\ref{dephase:eq:1})
along with the boundary conditions(continuity and the current
conservation at junctions J1 and J2) were solved to obtain the
amplitudes $t_u$, $t_d$, $r_u$ and $r_d$. Since the analytic
expressions are very lengthy, We confine ourselves to graphical
interpretation of the results. We have taken the flipper to be a
spin-half object ($M=2$) situated in the upper arm.  Now, depending
upon the initial state of the flipper we have possibility of either
spin-flip scattering ($\sigma_z=1/2,~S_z=-1/2$) or no spin-flip
scattering ($\sigma_z=1/2,~S_z=1/2$), as demanded by the conservation
of the total spin and its $z$-component. In the case of no-spin-flip
scattering ($\sigma_z=1/2,~S_z=1/2$) the problem at hand reduces to
that of simple potential scattering from the impurity.  We have set
$\hbar=2m=1$ and throughout the value of interaction strength
$G(=2mJ/\hbar ^2)$ is given in dimensionless units. The parameters
used for the analysis are mentioned in the figure captions.

\section{Results and discussion}

To begin with we first state the observed symmetry properties of the
transport coefficients in spin-flip scattering case where the electron
spin is opposite to the flipper spin. It is worth noting that due to
the presence of spin degree of freedom the problem in hand although
one-dimensional becomes a multi-channel problem. The spin-up reflection
coefficient $R_u=|r_u|^2$, spin-down reflection coefficient
$R_d=|r_d|^2$ and total reflection coefficient $R=R_u+R_d$ as a
function of the magnetic flux exhibit the AB-oscillations with flux
periodicity\cite{dephase:webb_ap} of $2\pi\phi_0$. All three
reflection coefficients are symmetric in the flux reversal as expected
on general grounds\cite{dephase:butt_ibm}.

\begin{figure*}[h]
\protect\centerline{\epsfxsize=4in \epsfbox{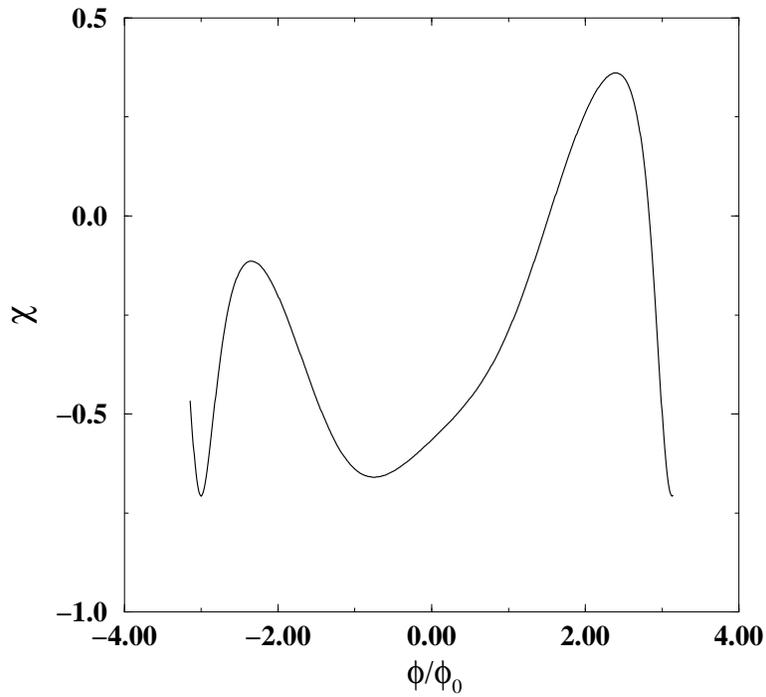}}
\caption{ Spin polarization ($\chi$) as a function of the flux $\phi$
  for interaction strength $G=10.0$. The lengths are $l_2/L=0.5$,
  $l_3/L = l_4/L = 0.25$ and $kL = 1.0$ }
\label{dephase:sppol}
\end{figure*}

The spin-up transmission coefficient
$T_u=|t_u|^2$ , spin-down transmission coefficient
$T_d=|t_d|^2$ which exhibit AB oscillations are asymmetric under flux reversal. The total transmission coefficient
$T=T_u+T_d$(related to the two-terminal
electrical conductance), however, is symmetric in flux reversal. 
 The transmission coefficient at flux $\phi$ for the
case when the incident particle is spin-up and the impurity is
spin-down is equal to the transmission coefficient for the case when
incident particle is spin-down and impurity is spin-up but the flux
direction is reversed.  For the spin-polarized transport the total
polarization $T_u-T_d$ is related to the spin-conductance
\cite{dephase:sarma}. The above symmetry properties imply that the
spin-conductance is asymmetric under the flux reversal. This can be
easily noted from Fig.~\ref{dephase:sppol}. In the figure we have plotted the
variation of spin polarization $\chi = (T_u-T_d)/T$ as a function of
the magnetic flux $\phi$. This spin-polarization can be experimentally
measured by using the well known spin-valve (magnetic valve or filter)
effect \cite{dephase:prinz}. It should be noted that at zero temperature the total
electrical and spin conductances are to be calculated by summing up
with equal weight-age the total transmission coefficients for all the
four cases, i.e., $\sigma_z=\pm 1/2$ and $S_z=\pm 1/2$.

\begin{figure*}[t]
\protect\centerline{\epsfxsize=4in \epsfbox{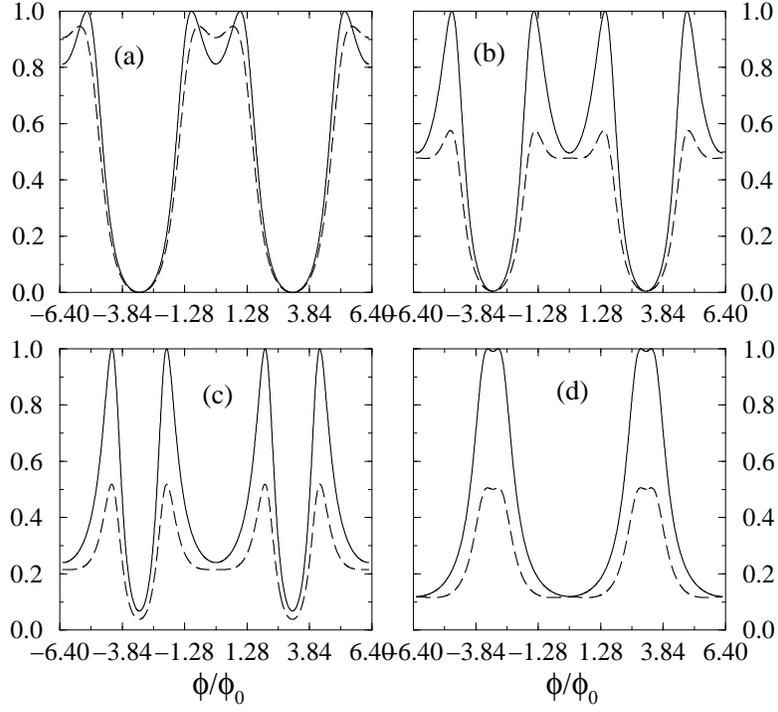}}
\caption{ Amplitude of AB oscillations or interference fringe visibility
  for the two cases of SFS and NSFS for different strengths of the
  exchange interaction. In all four cases $l_2/L=0.5$, $l_3/L = l_4/L
  = 0.25$ and $kL = 1.0$. The values of coupling strength $G$ are (a)
  $G = 1.0$, (b) $G = 5.0$, (c) $G = 10.0$ and (d) $G = 15.0$.}
\label{dephase:visib}
\end{figure*}

As discussed in the introduction, due to multiple reflections the
presence of a spin-flipper in one arm does not lead to "which-path"
information.  This would have implied the complete blocking of
spin-down transmission.  In contrast we clearly observe the
AB-oscillations for the case of $T_d$ originating from multiple
reflections. We now address the question of partial loss of
interference due to the spin-flipper. In Fig. \ref{dephase:visib} we have
plotted the total transmission coefficient $T=T_u+T_d$ for the
spin-flip scattering (SFS) case , and $T=T_u$ ($T_d=0$) for the no
spin-flip scattering (NSFS) case for different parameters as indicated
in the figures \ref{dephase:visib}(a-d).  As expected $T$ exhibits AB
oscillations which are periodic in flux with a period $2\pi\phi_0$ and
they are symmetric under flux reversal.  It is interesting to note,
however, that the interference fringe visibility (or the magnitude of
amplitude of AB oscillations) for the SFS case is always smaller than
that for the case of NSFS. This clearly indicates partial decoherence.

\begin{figure*}[h]
\protect\centerline{\epsfxsize=4in \epsfbox{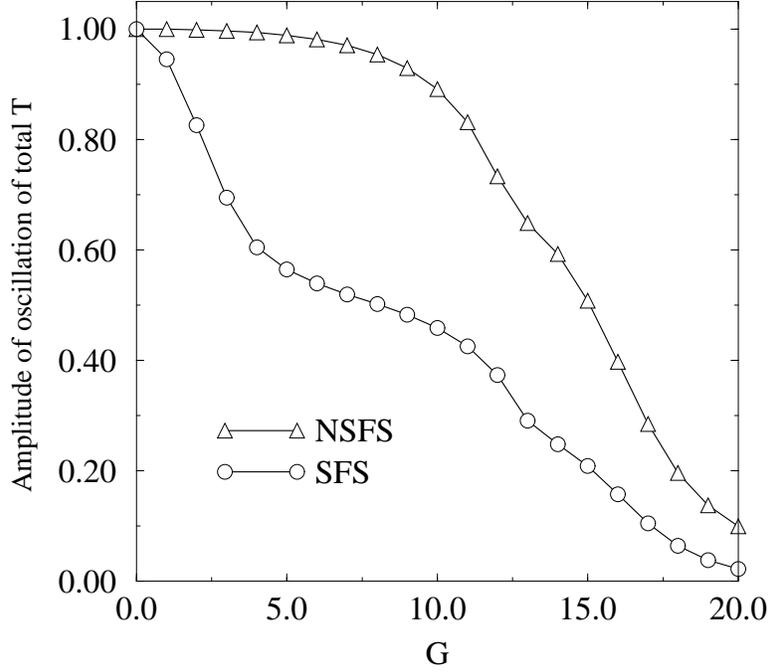}}
\caption{ Variation of Amplitude of AB oscillations with increasing
strength $G$ of spin-flipper for the case of asymmetrically placed
flipper. $l_3/L=0.15$, $l_4/L=0.35$ and $kL=1.0$.}
\label{dephase:asymm-amp}
\end{figure*}

To quantify the decoherence, we calculate the amplitude of AB
oscillations by taking the difference between the maximum and the
minimum of total transmission coefficient as a function of flux $\phi$
over one period of the oscillation. A plot of the variation of the
amplitude of oscillation of total transmission $T$ with the
interaction strength $G$ for the two cases, no spin-flip scattering
(NSFS: $S=1/2~m=1/2$) and spin-flip scattering (SFS: $S=1/2~m=-1/2$),
is shown in the figure Fig.~\ref{dephase:asymm-amp}. The signature of loss of
interference is that the amplitude of AB oscillation of transmission
coefficient for the spin-flip case is always smaller than that for the
no spin-flip case for all non-zero values of coupling strength $G$. In
other words the reduction of amplitude of AB oscillations is stronger
for the spin-flip scattering case. We have verified the above
observation for other parameters in the problem.  Thus the presence of
spin-flipper reduces the AB-oscillations. This substantiates our claim 
of decoherence due to entanglement.

Now We will turn our attention to current magnification and associated
effect of circulating currents as defined in earlier
works\cite{dephase:cme1,dephase:cme2}. Fig.~5 shows the plot of
circulating current density ($I_{c}$) versus $kL$ for the two separate
cases of spin-flip scattering and no-spin-flip scattering. When the
impurity spin is ``up'' the interaction does not allow spin-flip for a
spin-up incident electron and the impurity acts as a static potential
scatterer.  On the other hand when the impurity spin is ``down'' a
spin-flip scattering takes place. We compare the circulating current
densities for these two cases in order to see the role of entanglement
induced by the spin-flipper. The solid curve is for the no-flip case
while the dashed one is for the spin-flip case. The impurity strength
($G$) for both the cases is $4.0$.  In both the cases we take
$l_{2}/L=0.6$ and $l_{3}/L=l_{4}/L=0.2$. The figure shows that , the
circulating current for spin-flip case is significantly less than that
of the no-flip case in the range $12<kL<16$.  Thus one is led to
believe that the flipper acting as a dephasor suppresses the quantum
phenomena of current magnification.

However, this naive expectation turns out to be incorrect. This is
substantiated in Fig.~6 which shows circulating current densities for
the spin-flip and no-flip cases in the range $16<kL<19$ for the same
lengths as mentioned above. From this figure we see that in this
range of Fermi energies the amplitude of the circulating current is
actually enhanced in spite of the spin-flip scattering.

Thus the flipper can not only suppress the current magnification
effect but {\em can also} enhance it in some other range of Fermi
energies.  Thus far we have discussed how the flipper
affects current magnification
effect. The flipper also induces some new features. In Fig.~7 we have
plotted circulating current density ($I_{c}$)
versus $kL$ for $l_{3}/L=l_{4}/L=0.25$ and $l_{2}/L=0.5$ in the range
$5.6<kL<6.6$ shows an additional peak in the circulating current
density arising at a point corresponding to a minimum of spin-up
transmission (which is same as the maximum of the spin-down
transmission). This is indicative of the spin-flip process. This
effect is unique for the flipper having no counterpart in case of a
simple impurity, i.e., in this region ($5.6<kL<6.2$) no-flip
scattering case does not show any circulating current. This can be
ascribed to the additional phase shifts caused by spin-flip
scattering along-with multiple reflections. In the range $6.2<kL<6.6$
spin-flip scattering suppresses the current magnification.

Further, we see another interesting feature, namely the phenomenon of
current reversal. This is depicted in Fig.8. In this figure we plot the
circulating current density ($I_{c}$) versus $kL$ for $l_{3}/L = l_{4}/L =
0.3125$ and $l_{2}/L=0.375$ in the wave vector range $10<kL<15$ in which we
see that the spin-flip circulating current reverses its direction as
compared to the no-flip case, i.e., an anti-clockwise circulating current
for the no-flip case is converted into a clockwise one in the spin-flip
case.

In conclusion we have shown that presence of the spin-flipper which
reduces the AB oscillations(partial decoherence), need not
reduce the amplitude of
current magnification. In fact, in certain range of Fermi energies the
flipper enhances the current magnification. We believe that the
suppression of some quantum features and non-suppression of some other
quantum effects is a characteristic feature of entanglement,
environment consisting of finite degrees of freedom and the absence of
inelastic scattering. We expect the same to happen in other models
based only on the notion of entanglement.  Only the presence of
inelastic scattering(or coupling of a system to an environment with
infinite degrees of freedom), leading to irreversible loss of phase
memory, can dephase AB oscillations and reduce current magnification
simultaneously. Our analysis on the same model shows that two probe
spin-conductance
 is asymmetric in flux reversal as opposed to the
two probe electrical conductance which is symmetric. Further case of a
spin-flipper with higher number of internal states and that of
flippers in both arms of the ring are under investigation. We hope
that our results will stimulate further    interest and understanding
of dephasing and
decoherence arising from different models based on quantum
entanglement.

\section{Acknowledgements} 
The author thanks Sandeep K. Joshi, Dr. D. Sahoo and Colin Benjamin
for several useful discussions on this subject.

\vskip 3.0cm

\newpage

\begin{figure}
\protect\centerline{\epsfxsize=3.5in \epsfbox{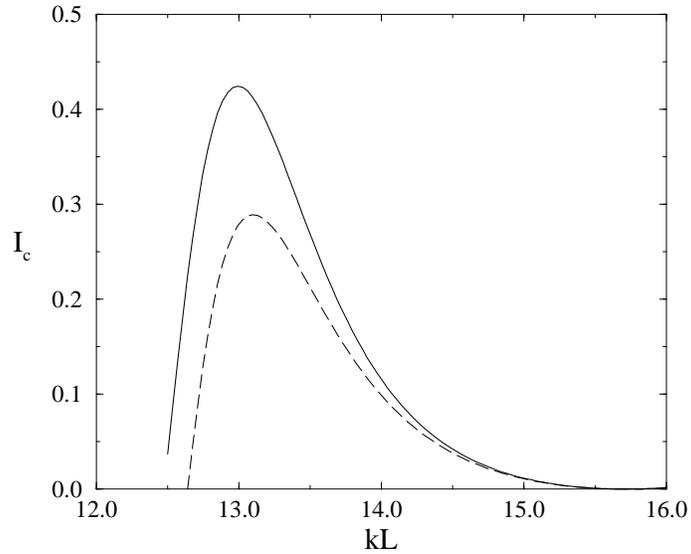}}
\caption{Plot of circulating current density $I_{c}$ versus $kL$. $G=4.0$ and
$ l_{2}/L = 0.6, l_{3}/L = l_{4}/L = 0.2 $ for both cases. 
The solid line is for the
no-flip case while the dashed line is for the spin-flip case.This
figure shows that the spin-flip process inhibits current
magnification.}
\label{fig1}
\end{figure}
 
\begin{figure}
\protect\centerline{\epsfxsize=3.5in \epsfbox{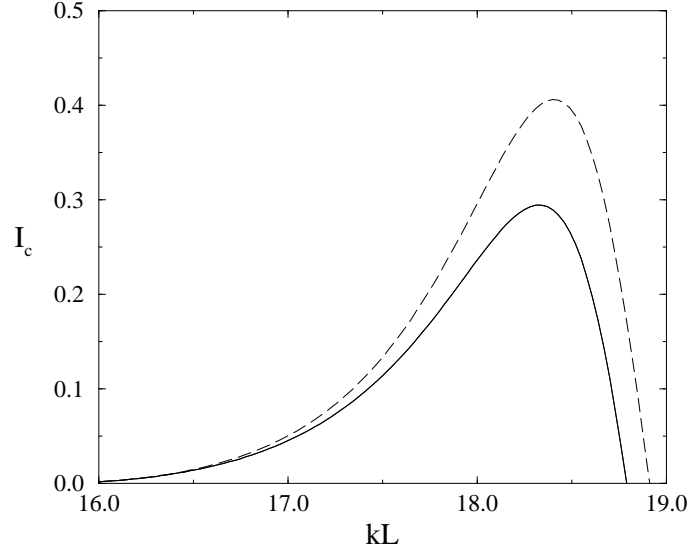}}
\caption{Plot of circulating current density $I_{c}$ versus $kL$. $G=4.0$ and
$ l_{2}/L = 0.6, l_{3}/L = l_{4}/L = 0.2 $ for both cases. 
The solid line is for the
no-flip case while the dashed line is for the spin-flip case. This
figure in contrast to Fig. 2 shows that the spin-flip process
enhances current magnification.}
\label{fig2}
\end{figure}

\begin{figure}
\protect\centerline{\epsfxsize=3.5in \epsfbox{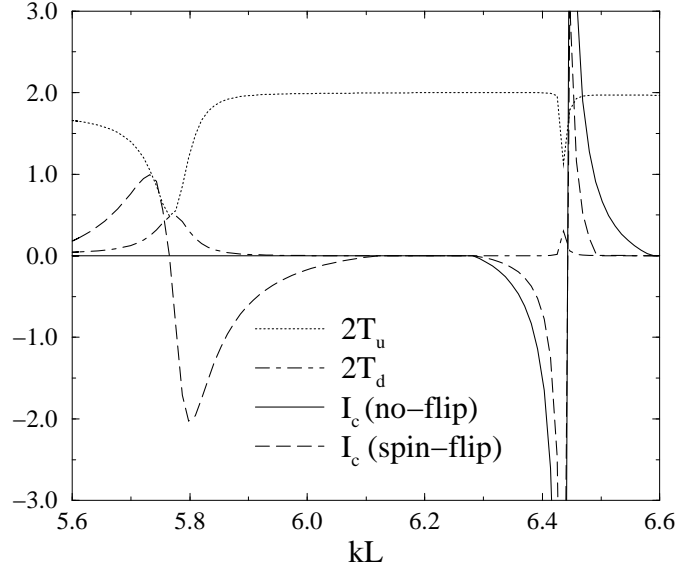}}
\caption{Plot of circulating current density $I_{c}$ versus $kL$. $G=4.0$ and
$ l_{2}/L = 0.5, l_{3}/L = l_{4}/L = 0.25 $ for both cases. The solid line
is for the no-flip case while the dashed line is for the spin-flip
case.The dash-dotted line is for $2T_{d}$ while the dotted line is
for $2T_{u}$ wherein $T_{u} = {\mid t_{u}\mid}^2$ and $T_{d} = {\mid
t_{d}\mid}^2$.}
\label{preso}
\end{figure}

\begin{figure}
\protect\centerline{\epsfxsize=3.5in \epsfbox{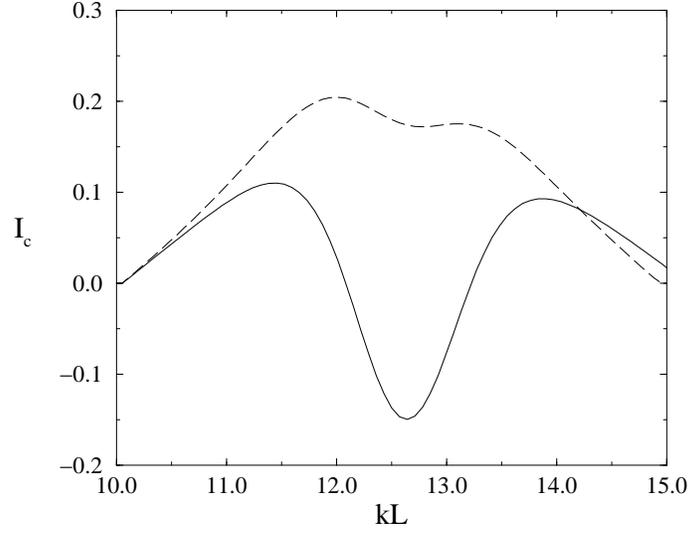}}
\caption{Plot of circulating current density $I_{c}$ versus $kL$. $G=4.0$ and
$l_{2}/L = 0.375, l_{3}/L = l_{4}/L = 0.3125$ for both cases. The solid
line is for the no-flip case while the dashed line is for the
spin-flip case.}
\label{prev}
\end{figure}

\end{document}